# Enhancement of anomalous Hall effect in Si/Fe multilayers

S. S. Das and M. Senthil Kumar[1]

*Department of Physics, Indian Institute of Technology Bombay, Mumbai 400 076, India*

## Abstract

Anomalous Hall effect studies were performed at 300 K on Si/Fe multilayers prepared by dc magnetron sputtering. About 60 times enhancement in the saturation Hall resistance and 80 times enhancement in anomalous Hall coefficient are obtained in $[Si(50Å)/Fe(t_{Fe})]_{20}$ multilayers when decreasing the Fe layer thickness from 100 Å to 20 Å. The largest anomalous Hall coefficient ($R_s$) of $1.4 \times 10^{-7}$ $\Omega$ m/T was found for $t_{Fe}$=20 Å, which is about three orders of magnitude larger than that of pure Fe and Fe/Cr, Al/Fe, Cu/Fe, $SiO_2$/FePt/$SiO_2$ multilayers. The ordinary Hall coefficient $R_0$ was about two orders of magnitude larger than that of pure Fe. The $R_s$ was found to vary with the longitudinal electronic resistivity, $\rho$ as $R_s \, \alpha \, \rho^{2.2}$, indicating the role of interfaces for the enhancement of the anomalous Hall effect in the multilayers. An increase of Hall sensitivity from 9 m$\Omega$/T to 1.2 $\Omega$/T is observed on decreasing $t_{Fe}$ from 100 Å to 10 Å. The high Hall sensitivity obtained is about three orders of magnitude larger than that of Al/Fe and Cu/Fe multilayers, showing it as an emerging candidate for Hall element for potential applications.

[1]Corresponding Author E-mail: senthil@iitb.ac.in

# 1. Introduction

Multilayer structures of different materials exhibiting interesting transport properties have brought the attention of the researchers for their rising demands for potential applications in modern technology. Since the discovery of giant magnetoresistance in Fe/Cr multilayers [1, 2], multilayers of various combinations of magnetic layers separated by a non magnetic conducting spacer are being studied extensively. Recently, peculiar magnetic and transport properties have been reported by many authors with the introduction of a non-magnetic semiconducting spacer (Si, Ge, GaAs, ZnTe etc.) in place of a conducting one [3]. Anomalous Hall effect (AHE) which is an electronic transport phenomenon dependent on spin-orbit interaction found in magnetic materials emerges as a sensitive tool for magnetic characterization of nanostructures, thin films, recording layer of double-layered perpendicular magnetic recording media, where conventional vibrating sample magnetometer fails [4, 5]. High resistivity, low frequency response and strong temperature dependence, complicated synthesis process which a semiconducting sensor based on ordinary Hall effect faces, AHE shows a new approach for magnetic sensing. High sensitivity, weak temperature dependence, linear field response, hysteresis free behavior with field and of course low manufacture cost of the AHE based devices, open up new possibilities in the modern field of sensors.

Although Hall effect has been discovered more than a century ago, the smaller Hall coefficient and Hall sensitivity values of bulk magnetic materials has limited its application possibilities in the field of spintronics. Recently, lots of researches have been devoted to enhance the Hall coefficient and Hall sensitivities so as to make the effect suitable for modern technological applications. Attempts have been made to study the effect in the case of magnetic multilayers and thin films. Lu et al. have reported the highest anomalous Hall sensitivity of 12000 Ω/T in $SiO_2$/FePt/$SiO_2$ sandwich structure films which is about an order of magnitude larger than the semiconductor sensitivity [6]. Accordingly, AHE in various multilayer structures like Fe/Cr [7, 8], Fe/Ge [9], CoFe/Pt [10] have been reported by many authors but so far no reports on Si/Fe multilayers. In this paper, we have investigated the effect of introducing an intrinsic Si spacer on the AHE of the Si/Fe multilayers.

## 2. Experimental details

The Si/Fe multilayers were deposited onto glass substrates by dc magnetron sputtering at ambient temperature. An argon pressure of 0.004 mbar was used during sputtering and a base pressure of $2 \times 10^{-6}$ mbar was achieved prior to the depositions. Sputtering power was kept fixed at 40 W for both Fe and Si targets. The multilayer structure was obtained by exposing the substrates placed on a computer controlled rotating substrate holder to Si and Fe targets alternately. Two series of samples were prepared, one of the form [Si(50 Å)/Fe($t_{Fe}$)]$_{20}$ and the other of the form [Si($t_{Si}$)/Fe(20 Å)]$_{20}$, where $t_{Fe}$ and $t_{Si}$ are the nominal thickness values of Fe and Si layers, respectively. In the first series, $t_{Fe}$ was varied from 10 to 100 Å whereas for the second series the $t_{Si}$ varied from 10 to 80 Å. Multilayers and single layers of the form [Al(50 Å)/Fe(20 Å)]$_{20}$, [Cu(50 Å)/Fe(20 Å)]$_{20}$, [Si(50 Å)/Al(20 Å)]$_{20}$, [Si(50 Å)/Cu(20 Å)]$_{20}$, Al(1000 Å), Cu(1000 Å) were also prepared under similar conditions for comparison. The structural and microstructural analysis of the samples were done by high resolution transmission electron microscopy (HRTEM) using JEOL JEM 2100F instrument operating at an accelerating voltage of 200 kV. Magnetization measurements at 300K were performed by using a vibrating sample magnetometer, an attachment of a Physical Property Measurement System of Quantum design, Inc. with the magnetic field applied up to 90 kOe in the film plane. The Hall effect measurements were carried out at 300K by circular four probe method by applying magnetic field perpendicular to the film plane up to 27 kOe.

## 3. Results and discussions

### 3.1. Anomalous Hall effect in [Si(50 Å)/Fe(t$_{Fe}$)]$_{20}$ multilayers

In figure 1, we have plotted the Hall resistance (R$_h$) versus magnetic field (H) for [Si(50 Å)/Fe(t$_{Fe}$)]$_{20}$ multilayers measured at 300K where t$_{Fe}$ was varied in the range 10-100 Å. A strong increase in the Hall signal with decreasing t$_{Fe}$ is clearly seen from the figure. The expression for Hall resistance can be written as [11, 9, 12-14]

$$R_h = \frac{\rho_h}{t} = \frac{R_0 H + R_s 4\pi M}{t} \tag{1}$$

where $\rho_h$, $t$, $R_0$, $R_s$ and $M$ are Hall resistivity, thickness of the entire multilayer stack of the samples, ordinary Hall coefficient, anomalous Hall coefficient and magnetization, respectively. The saturated anomalous Hall resistance ($R_{hs}^A$) can be calculated by subtracting the ordinary Hall term from the linear fit of the high field regime Hall data where the anomalous Hall term saturates. The values of R$_0$, R$_s$ and perpendicular saturation field ($H_s = 4\pi M_s$) can be calculated from the Hall effect data following the procedure as described in refs. 7, 14. As seen from figure 1, in the low field regime (–9 kOe to +9 kOe) the Hall resistance has nearly linear dependence with the field.

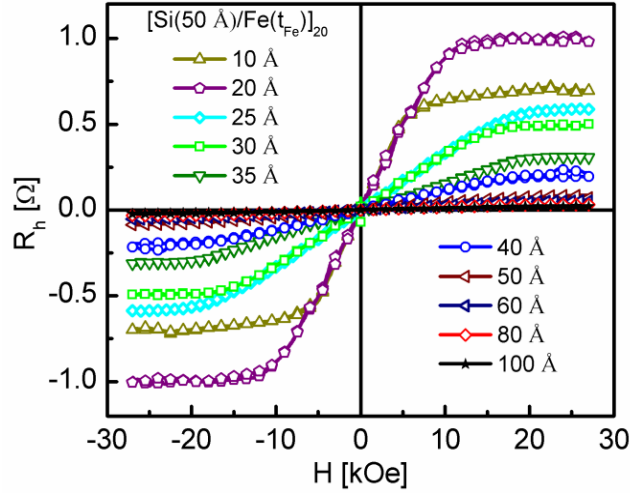

Figure 1. The Hall resistance ($R_h$) versus magnetic field (H) for the [Si(50 Å)/Fe($t_{Fe}$)]$_{20}$ multilayers at 300 K. Here $t_{Fe}$ is varied between 10 Å to 100 Å.

Figure 2 shows the $t_{Fe}$ dependence of the $R_{hs}^A$ and $R_s$ normalized with respect to that of $t_{Fe}$=100 Å. As can be seen from the figure, $R_{hs}^A$ for the sample with $t_{Fe}$=20 Å is about 60 times enhanced when compared with that for $t_{Fe}$=100 Å. Similarly, about 80 times enhancement is observed for $R_s$. The large enhancement observed by us is much larger than that reported in the case of single Fe and Ni films. An increase of $R_{hs}^A$ of about 6-8 times and $R_s$ of about 4 times has been reported in the case of Ni single film when decreasing the thickness from 200 Å to 20 Å [15]. In the case of Fe single films, about 3-4 times increase in $R_{hs}^A$ and $R_s$ is reported when the film thickness is reduced from 1700 Å to 80 Å [16]. About 3-5 times increase in $R_{hs}^A$ and $R_s$ values has also been reported in the case of Fe/Cr, Fe/Cu and Fe/Ge multilayers [7, 9, 17]. In the case of SiO$_2$/FePt/SiO$_2$ sandwich structure which shows highest anomalous Hall sensitivity till now, a change of about 2 times in the AHE was reported within the thickness range of 14 Å to 200 Å [6]. When compared with the reported value of 0.0002 Ω for bulk Fe [18], the $R_{hs}^{A}$ we have observed is about 4 orders of magnitude larger. Similarly, the largest value of $R_s$ i.e. $1.4 \times 10^{-7}$ Ω m/T observed in Si/Fe multilayers for $t_{Fe}$ = 20 Å is about 3 orders of magnitude larger than that reported for bulk Fe [9, 13] i.e. $(2.8-7.2) \times 10^{-10}$ Ω m/T. Because of the large enhancements in the AHE as compared to the homogeneous ferromagnets, some authors also called it as "Giant Hall effect" as an analogy to the giant magnetoresistance [18, 19]. However, in the case of the Si/Fe multilayers the giant enhancement in the AHE observed is not just because of the simple roughness and interdiffusion effects like in other multilayers but also due to a metal-semiconductor interface where the magneto-transport is more complex due to a mixed metal-semiconducting behavior of the charge carriers. When compared with the Fe/Ge multilayers [9], the enhancement of AHE in our Si/Fe multilayers is quite large. This is because of Ge has higher conductivity of 2.2 $Ω^{-1}$ $m^{-1}$ as compared to that of Si $4 \times 10^{-4}$ $Ω^{-1}$ $m^{-1}$ at 300 K [20]. Besides this, the charge carriers in the Fe/Ge multilayers [9] are electrons whereas that in the Si/Fe multilayers are holes. At 300 K, the mobility of electrons in Ge is 0.38 $m^2$/V-s whereas that of holes in Si is 0.05 $m^2$/V-s [20]. This higher conductivity and carrier mobility in Ge as compared to that of Si increases the possibility of short-circuit and shunting effects in the Fe/Ge multilayers and thus could be responsible for its decrease in the AHE signal. Thus, the large enhancement observed by us in $R_{hs}^{A}$ and $R_s$ is much larger than that reported in any material so far. The large enhancement in $R_{hs}^{A}$ and $R_s$ in our samples upon decreasing $t_{Fe}$ may be due to the interface scattering resulting from the interdiffusion, interface roughness and interface alloy formation. In our samples, $R_{hs}^{A}$ and $R_s$ were found to decrease for $t_{Fe}$=10 Å because of the loss of ferromagnetism due to the formation of superparamagnetic fine Fe grains. Our zero field cooled and field cooled magnetization versus temperature data (not shown) also shows the appearance superparamagnetism for $t_{Fe}$=10 Å. This suggests that there exists a critical value of $t_{Fe}$ at which $R_{hs}^{A}$ and $R_s$ reach their maximum.

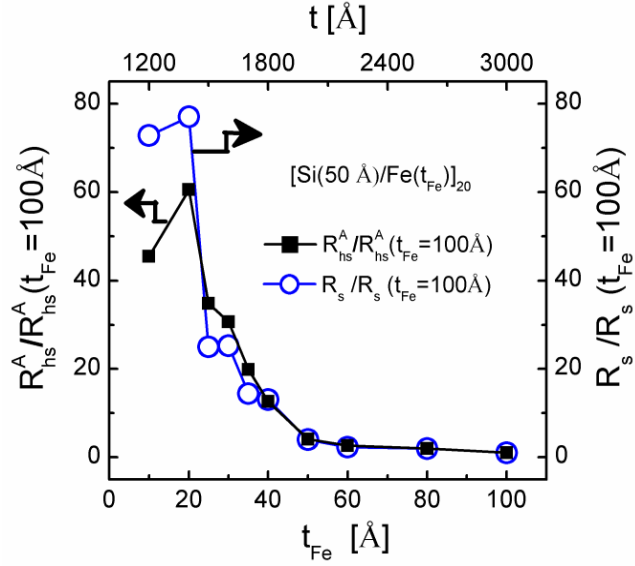

Figure 2. The $t_{Fe}$ dependence of $R_{hs}^{A}$ and $R_s$ normalized with respect to that of $t_{Fe}$=100 Å. Total thickness t of the multilayer stack is plotted in the top x-axis. Large changes of about 60 and 80 times have been observed in $R_{hs}^{A}$ and $R_s$, respectively.

To support our explanation of the enhancement of AHE due to interface effects, we have performed structural and microstructural studies on the samples through X-ray diffraction analysis (XRD) [21] and high resolution transmission electron microscopy (HR-TEM) on the samples. The XRD data of the Si/Fe multilayers which we have already reported show a decrease of grain size from about 100 Å to 30 Å with the decrease of $t_{Fe}$ from 100 Å to 30 Å [21]. The grain sizes for the samples with $t_{Fe}$ less than 30 Å could not be obtained due to the absence of the diffraction peaks in the XRD patterns. The decrease of the grain size with decreasing $t_{Fe}$ results in an increase in surface to volume ratio thus enhancing the surface and interface effects. Using HR-TEM, the selected area electron diffraction pattern obtained for the [Si(50Å)/Fe(20Å)]$_4$ multilayers is shown in the inset of figure 3(a). It shows the rings corresponding to the Fe layers only and there is no diffraction pattern from the Si layers. This suggests that the Fe layers are nanocrystalline whereas the Si layers are amorphous in nature. Similar features have also been observed from the XRD data of our samples [21]. These behaviours are also consistent with the results reported by some authors [22, 23]. Figure 3(a) and (b) represent the HR-TEM images of [Si(50Å)/Fe(20Å)]$_4$ and [Si(50Å)/Fe(50Å)]$_4$ multilayers, respectively. As can be seen from these figures, the separation between the Fe grains is large in [Si(50Å)/Fe(20Å)]$_4$ multilayers as compared to that of the [Si(50Å)/Fe(50Å)]$_4$ multilayers suggesting that the former is more discontinuous than the latter. Therefore, with the decrease of $t_{Fe}$ the continuity of the Fe layer decreases which leads to high surface and interface scattering in the Si/Fe multilayers. This scattering in turn causes the enhancement of AHE and longitudinal electronic resistivity of the multilayers.

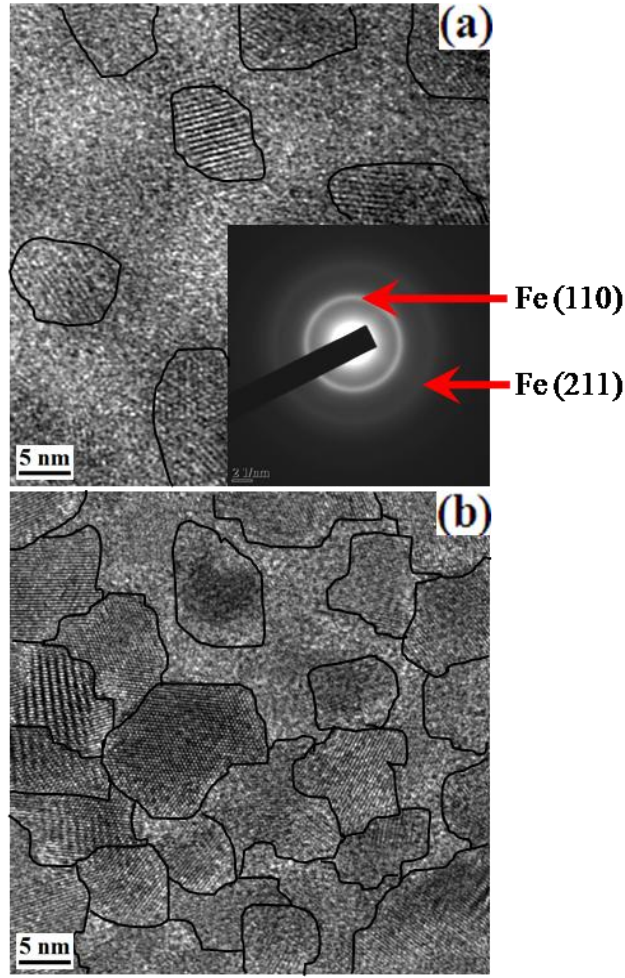

Figure 3. (a) HR-TEM image of [Si(50 Å)/Fe(20 Å)]$_4$ multilayers. The solid curves are drawn to show the grains. The inset shows the selected area electron diffraction pattern of this sample. (b) HR-TEM image of [Si(50 Å)/Fe(50 Å)]$_4$ multilayers.

As the AHE mainly depends on the magnetization of the sample we have performed magnetization measurements on the samples at 300 K with an in-plane magnetic field upto 90 kOe. The magnetization data of the samples show soft ferromagnetic behavior with strong in-plane magnetic anisotropy. Figure 4(a) shows the magnetization curves of selected samples of the [Si(50 Å)/Fe($t_{Fe}$)]$_{20}$ multilayers. The inset of the figure 4(a) shows the coercivity of ~20 Oe for the [Si(50 Å)/Fe(20 Å)]$_{20}$ multilayers. The low coercivity of the multilayers is an indication of the presence of Fe nanograins which is also observed from the XRD and HR-TEM data. The saturation magnetization ($M_s$) obtained from the magnetization data of all the samples is plotted as a function of $t_{Fe}$ in figure 4(b). The $M_s$ is almost constant around 1550 emu/cc for larger $t_{Fe}$ and it decreases with the decrease of $t_{Fe}$. The bulk value of $M_s$ of Fe is 1714 emu/cc [24]. Such a decrease of $M_s$ with the decrease of $t_{Fe}$ (below 50 Å) is commonly observed in magnetic thin films. Our magnetization data shows the decrease of $M_s$ with the decrease of $t_{Fe}$ whereas the AHE which mainly depends on the magnetization of the material strongly increases with the decrease of $t_{Fe}$. Thus, the enhancement of the AHE in our Si/Fe multilayers indicates that the surface/interface effects are much more dominant than the reduction in $M_s$.

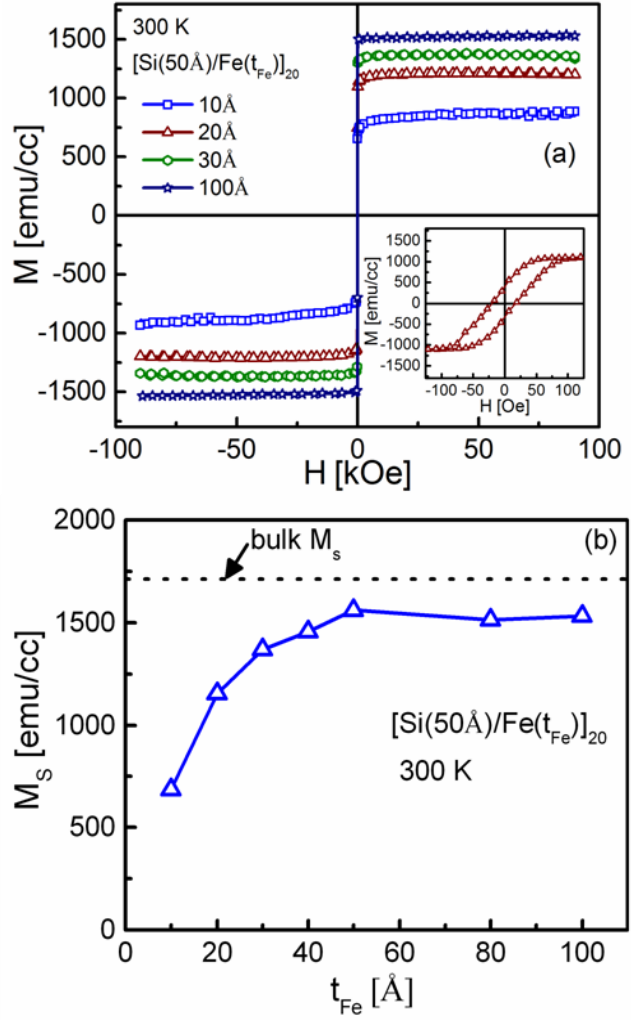

Figure 4. (a) In-plane magnetization loops of selected samples of the [Si(50 Å)/Fe($t_{Fe}$)]$_{20}$ multilayers at 300 K. The inset shows the same loop in an expanded x-scale for $t_{Fe}$ = 20 Å which shows the highest AHE. (b) Variation of saturation magnetization ($M_s$) of the [Si(50 Å)/Fe($t_{Fe}$)]$_{20}$ multilayers with $t_{Fe}$ at 300 K. The open triangles represent the experimental data points and the solid line is the guide to eyes. The dashed line shows the $M_s$ of bulk Fe at 300 K i.e. $M_s^{bulk}$=1714emu/cc.

## 3.2. Anomalous Hall effect in [Si($t_{Si}$)/Fe(20 Å)]$_{20}$ multilayers

To understand the effect of Si spacer layer on the Hall resistance of the Si/Fe multilayers, we also have performed Hall effect measurements at 300 K by varying the Si layer thickness ($t_{Si}$) and keeping the Fe layer thickness fixed at 20 Å which corresponds to maximum value of $R_{hs}^A$ in figure 2. The $R_h$ versus H graph of [Si($t_{Si}$)/Fe(20 Å)]$_{20}$ multilayers is shown in the figure 5(a). The $t_{Si}$ dependence of $R_{hs}^A$ and $R_s$ normalized with respect to that of $t_{Si}$=10 Å is shown in the figure 5(b). Our data suggests that $R_{hs}^A$ and $R_s$ increase with $t_{Si}$ and shows maximum at $t_{Si}$=50 Å and then start decreasing on further increasing $t_{Si}$. Thus, similar to the dependence on $t_{Fe}$, there also exists a critical value of $t_{Si}$ for which $R_{hs}^A$ and $R_s$ show their maximum. Therefore, similar to classical percolation threshold in granular metal insulator systems like FeSiO$_2$ [19], CoSiO$_2$ [25], NiFeSiO$_2$ [26] and critical bilayer period in multilayer systems like Fe/Ge [9], there also exists critical values of $t_{Fe}$ and $t_{Si}$ in Si/Fe multilayers for which Hall resistance becomes maximum. The critical value of $t_{Fe}$ found for [Si(50 Å)/Fe($t_{Fe}$)]$_{20}$ multilayers is 20 Å and that of $t_{Si}$ for [Si($t_{Si}$)/Fe(20Å)]$_{20}$ multilayers is 50 Å.

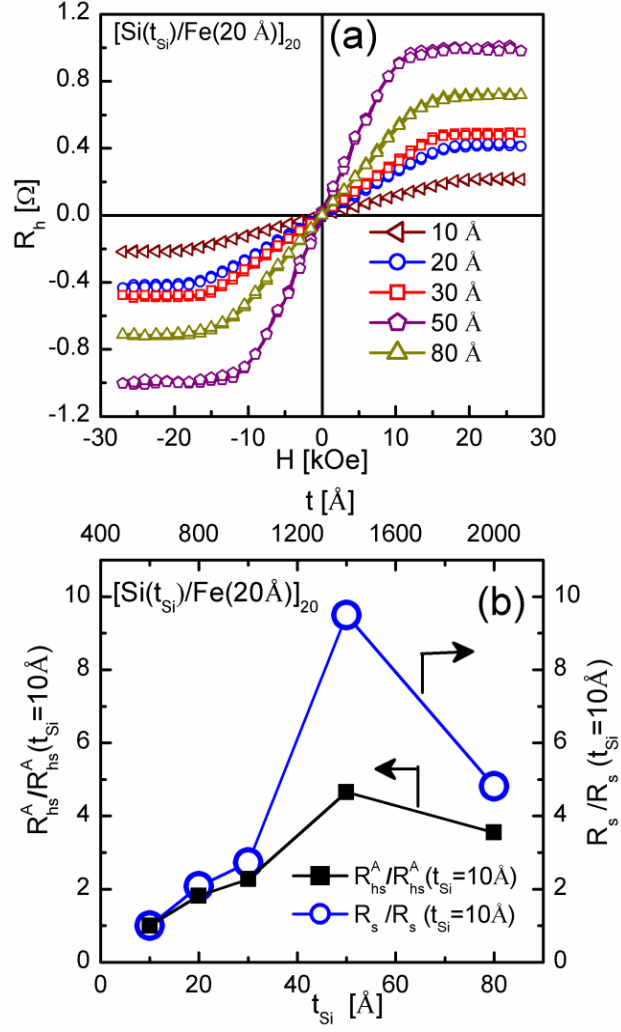

Figure 5. (a) The Hall resistance ($R_h$) versus magnetic field (H) for $[Si(t_{Si})/Fe(20\text{Å})]_{20}$ multilayers at 300 K. (b) The $t_{Si}$ dependence of $R_{hs}^A$ and $R_s$ normalized with respect to that of $t_{Si}$=10 Å. Total thickness t of the multilayer stack is plotted in top x-axis.

### 3.3. Verification of Hall effect due to Si layers

To verify the individual contributions of Fe and Si layers in the Si/Fe multilayers to the total Hall resistance and the sign of the Hall coefficients, measurements were performed on $[Si(50\text{ Å})/Al(20\text{ Å})]_{20}$, $[Si(50\text{ Å})/Cu(20\text{ Å})]_{20}$, $[Al(50\text{Å})/Fe(20\text{Å})]_{20}$ and $[Cu(50\text{Å})/Fe(20\text{Å})]_{20}$ multilayers and also on Al (1000Å) and Cu (1000Å) single layers prepared under identical conditions. We have chosen Al and Cu spacer layers which are nonmagnetic metals generally used in silicon integrated circuits [27, 28]. In figure 6(a) the $R_h$ data obtained for the above mentioned samples along with the data of [Si(50 Å)/Fe(20 Å)]$_{20}$ are plotted. The data points other than for [Si(50Å)/Fe(20Å)]$_{20}$ lie very close to each other. It is clearly seen from the figure that the signals obtained from these samples are much smaller than that for the [Si(50 Å)/Fe(20 Å)]$_{20}$. In figure 6(b), the same data shown in figure 6(a) for some samples are shown in an expanded y-scale. The $R_h$ values of the Si/Al and Si/Cu multilayers are approximately of same magnitudes as that of Al and Cu single layers respectively, indicating that the Si spacer layer has negligible contribution to the total Hall resistance. Thus, the Hall signal in Si/Fe multilayers basically comes from the magnetic Fe layers and from the interfaces. Figure 6(c) shows the $R_h$ versus H plot for [Al(50Å)/Fe(20Å)]$_{20}$ and [Cu(50Å)/Fe(20Å)]$_{20}$ multilayers. Our Hall effect measurements show that the $R_{hs}^A$ value for Si/Fe multilayers is more than two orders of magnitude larger than that of Al/Fe and Cu/Fe multilayers. Furthermore, the introduction of a semiconducting Si spacer instead of conducting Al or Cu spacer enhances the Hall signal in Si/Fe multilayers by reducing the short circuit and shunting effects. Xu et al. have reported that anomalous Hall effect in Fe/Cu multilayers is decreased due to short circuit and shunting effects caused by conducting Cu layers [17]. This clearly shows how Si spacer layer play a major role in the enhancement of AHE by improving the interface effects. Therefore, by optimizing $t_{Fe}$ and $t_{Si}$ of the multilayer stacks the short circuit and shunting effects can be minimized.

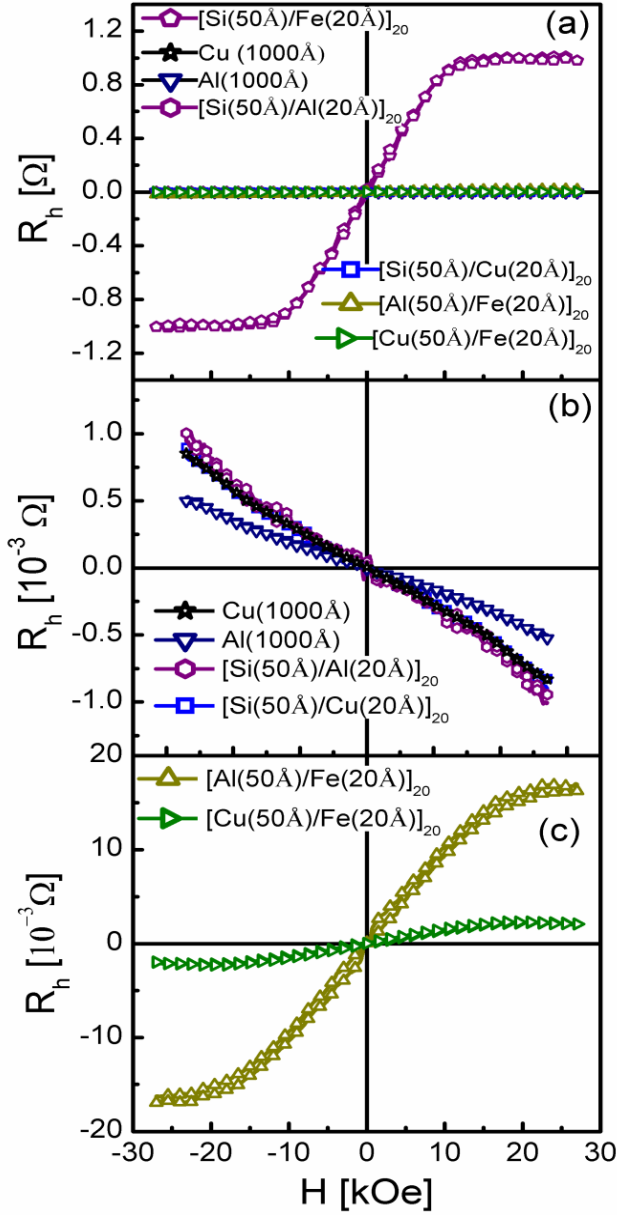

Figure 6. (a) The $R_h$ versus H graph of $[Si(50Å)/Fe(20Å)]_{20}$, $[Si(50Å)/Al(20Å)]_{20}$, $[Si(50Å)/Cu(20Å)]_{20}$, $[Al(50Å)/Fe(20Å)]_{20}$, $[Cu(50Å)/Fe(20Å)]_{20}$ multilayers, Al (1000Å) and Cu (1000Å) single layer samples. The data points other than for $[Si(50Å)/Fe(20Å)]_{20}$ lie very close to each other. (b) The same $R_h$ versus H data shown above is plotted in an expanded y-scale to show the data for $[Si(50Å)/Al(20Å)]_{20}$, $[Si(50Å)/Cu(20Å)]_{20}$ multilayers, Al (1000Å) and Cu

(1000Å) single layers. (c) The same $R_h$ versus H data shown above is plotted in an expanded y-scale to show the data for [Al(50Å)/Fe(20Å)]$_{20}$ and [Cu(50Å)/Fe(20Å)]$_{20}$ multilayers.

### 3.4. Ordinary Hall effect

The Ordinary Hall effect contribution was extracted from the high field regime Hall effect data of the Si/Fe multilayers. The normal Hall coefficient ($R_0$) as computed from the linear fit of the high field linear regime Hall effect data for the [Si(50Å)/Fe($t_{Fe}$)]$_{20}$ multilayers is plotted as a function of $t_{Fe}$ as shown in the figure 7. Similar to pure Fe [13], the sign of $R_0$ is also found to be positive indicating the conduction mechanism in the Si/Fe multilayers is due to positive charge carriers. Upon decreasing $t_{Fe}$ from 100 Å to 20 Å about 10 times increase of $R_0$ is observed. Similar trend of $R_0$ with the film thickness has also been reported in the case of pure Fe [16]. The value of $R_0$ obtained for our multilayers with $t_{Fe}$ = 20 Å is 8.8 × 10$^{-9}$ Ω m/T. This value of $R_0$ is about two orders of magnitude larger than that of pure Fe (~ 10$^{-11}$ Ω m/T) [18, 13]. We have also calculated the carrier concentration ($n'$) using the relation, $R_0 = 1/n'e$, where e = 1.6 × 10$^{-19}$ C. The calculated $n'$ is of the order of 10$^{22}$ cm$^{-3}$ for the samples with $t_{Fe}$ > 20 Å and for the samples with $t_{Fe}$ = 10 and 20 Å, $n'$ value was about 10$^{21}$ cm$^{-3}$. The $n'$ value as reported in the case of pure Fe is of the order of 10$^{23}$ cm$^{-3}$ whereas that for Fe/Ge multilayers is of the order of 10$^{21}$ cm$^{-3}$ [9]. The decrease of the $n'$ in our samples upon decreasing $t_{Fe}$ may be due to the interfaces. Thus, the interface scattering effects in our samples results in the large enhancement of about 10 times in $R_0$ which has only nonmagnetic scattering part originating due to the Lorentz force on the charge carriers. On the other hand, the interface scattering due to the magnetic contribution causes even larger enhancement of about 80 times in the case of $R_s$ as discussed in section 3.1.

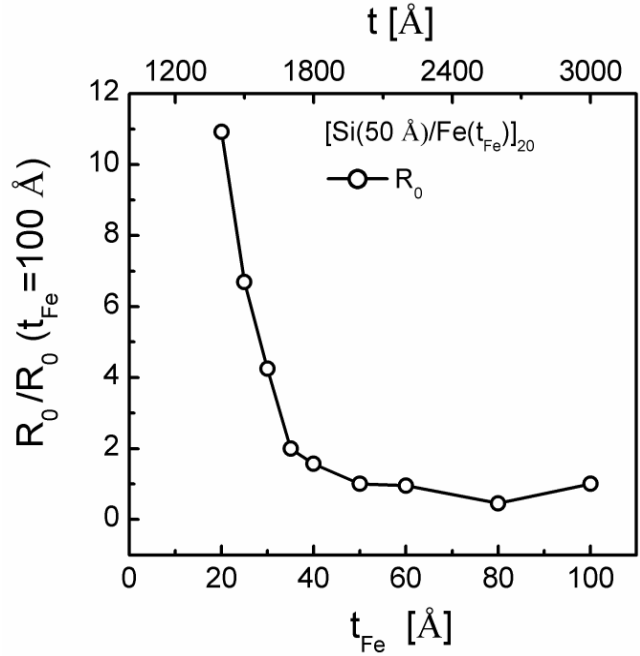

Figure 7. The normal Hall coefficient $R_0$ versus $t_{Fe}$ for the [Si(50 Å)/Fe($t_{Fe}$)]$_{20}$ multilayers. Total thickness t of the multilayer stack is plotted in the top x-axis. About 10 times increase in $R_0$ is observed.

### 3.5. Mechanism of AHE in the [Si(50 Å)/Fe($t_{Fe}$)]$_{20}$ multilayers

In figure 8, we have analyzed the mechanism responsible for AHE in our samples using scaling law. The mechanism of anomalous Hall effect as purposed by Karplus and Luttinger [29] is due to the spin-orbit interaction of polarized conduction electrons and is intrinsic in nature. They suggested that the AHE has a quadratic dependence with longitudinal electronic resistivity ($\rho$). Later Smit [30] argued that anomalous Hall effect being extrinsic in nature should be due to the break of right left symmetry due to spin orbit scattering of conduction electrons on impurities and/or phonons. On the basis of extrinsic spin orbit scattering, two mechanisms named classical skew scattering and nonclassical side-jump have been purposed by Smit [30] and Berger [31]. According to skew-scattering and side jump models the relation between anomalous Hall coefficient ($R_s$) and longitudinal electronic resistivity, $\rho$ can be expressed theoretically as

$$R_s \propto \rho^n \, (\text{or} \, \rho_h \propto \rho^n), \qquad (2)$$

where n = 1 and 2 represents the skew scattering and side jump mechanisms, respectively. In most of the low resistive ferromagnetic materials, the n value closed to 1 has been reported whereas in the case of high resistive ferromagnetic materials, n value closed to 2 was reported [32]. However any additional contribution due to phonon scattering, surface-interface effects etc. can sometimes results in the n value even larger than 2. In general, equation (2) which is also called as scaling law can be used for fitting of the experimental data and the value of 'n' thus obtained will give the dominant mechanism responsible for AHE.

To understand the mechanism responsible for the AHE in Si/Fe multilayers, we have verified the scaling law between $R_s$ and $\rho$ using equation (2) by plotting the ln $R_s$ versus ln $\rho$ as shown in the figure 8. Here $\rho$ is the longitudinal electronic resistivity in zero applied magnetic field. The value of the exponent 'n' thus obtained from the above linear fitting of ln $R_s$ versus ln $\rho$ is 2.2 ± 0.1 which suggests that the side jump mechanism is the dominant mechanism for the observed AHE in Si/Fe multilayers. This large value of 'n' suggests the role of interfaces in the enhancement of AHE in the multilayer structures. The 'n' value for sputtered Fe/Ge multilayers as reported by Liu et al. was 1, suggesting that the skew scattering is the dominant mechanism in their samples [9]. Similarly, Song et al. have found n = 2.6 in Fe/Cr multilayers deposited by electron beam evaporation [7]. They claimed that spin dependent interface scattering may give rise to the large values of 'n' and hence the side jump mechanism is responsible for AHE in their system. The value of 'n' in the case of molecular beam epitaxy grown Co/Cu multilayers as reported by Tsui et al. was 2 suggesting the side jump mechanism as the dominant one [33]. They claimed that the connection between the bulk-like scattering which mainly depends on Co magnetization and the interface scattering leads to the value of n = 2. Guo et al. have studied the Co/Pd multilayer system and found surprisingly large value of n = 5.7 [34]. They suggested that the interface scattering plays a major role in heterogeneous systems for the large value of n. Very different values of n greater than 2 in those systems suggest that the scaling law which is developed for homogeneous ferromagnets may not hold well in the case of heterogeneous ferromagnetic systems. The theoretical study of the AHE in multilayer structures by Zhang suggests that the n=2 is the case when the mean free path of the electrons is less than the thickness of the layers [17, 35]. In our case, the large value n = 2.2 obtained from the above scaling law may also be due to the significant contribution from the interface effects arising because of interface roughness, interdiffusion and interface alloy formation which were not included in the scaling law of AHE.

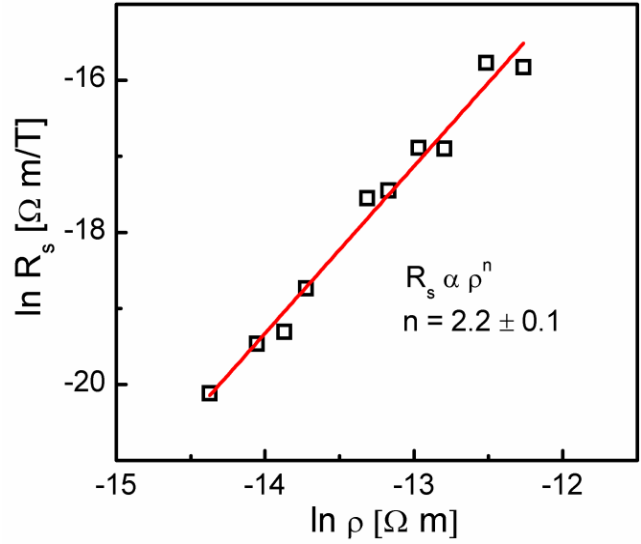

Figure 8. The ln $R_s$ versus ln $\rho$ plot to verify the scaling law between $R_s$ and $\rho$ for [Si(50 Å)/Fe($t_{Fe}$)]$_{20}$ multilayers. Here $\rho$ value was measured at zero applied magnetic field. The open squares are the data points and the solid line is the linear fit according to equation. (2), i.e. $R_s \propto \rho^n$.

## 3.6. Hall sensitivity

The Hall sensitivity (S) defined by $dR_h/dH$ is plotted as a function of $t_{Fe}$ in figure 9 for [Si(50Å)/Fe($t_{Fe}$)]$_{20}$ multilayers in the low field range where $R_h$ linearly depends on H. An increase of the Hall sensitivity from 9 mΩ/T to 1.2 Ω/T has been observed on decreasing $t_{Fe}$ from 100 Å to 10 Å. The samples with $t_{Fe}$=10 Å and 20 Å show the Hall sensitivities of 1.2 Ω/T and 0.9 Ω/T which are about three orders of magnitude larger than that of Al/Fe and Cu/Fe multilayers prepared under similar conditions. A Hall sensitivity of 1.0 Ω/T has also been reported for the Fe/Ge multilayers [9]. Recently, a large anomalous Hall sensitivity of about 12000 Ω/T has been reported in the case of SiO$_2$/FePt/SiO$_2$ sandwich films where perpendicular anisotropy of the FePt causes this high value of sensitivity [6]. The increase in the Hall sensitivity with the decrease of $t_{Fe}$ in our Si/Fe multilayers having in-plane magnetic anisotropy is a consequence of the increase of the anomalous Hall effect due to surface and interface scattering.

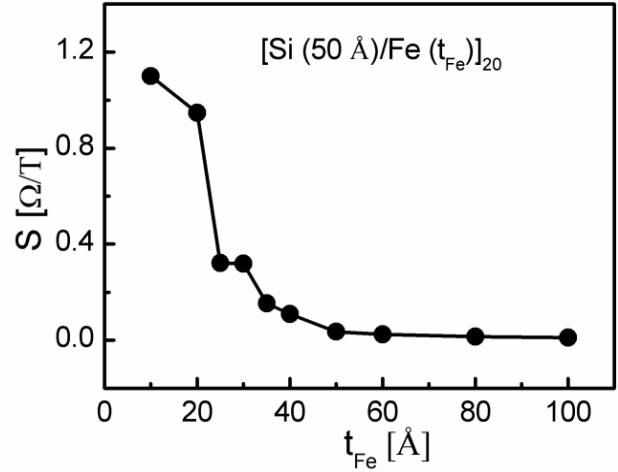

Figure 9. The Hall sensitivity (S) of $[Si(50Å)/Fe(t_{Fe})]_{20}$ multilayers plotted as a function of $t_{Fe}$.

## 4. Conclusion

In conclusion, we have observed about 60 times enhancement of $R_{hs}^{A}$ and 80 times enhancement of $R_s$ in $[Si(50 Å)/Fe(t_{Fe})]_{20}$ multilayers upon reducing the thickness of the Fe layers. Measurements of AHE in Si/Al, Si/Cu, Fe/Al, Fe/Cu multilayers confirmed that this enhancement is due to the combination of Si and Fe in multilayer form. In addition, there exist critical values of $t_{Fe}$ and $t_{Si}$ for which Hall resistance reaches its maximum. The $R_0$ value for $t_{Fe}$ = 20 Å was about 2 orders of magnitude larger than that of pure Fe. The scaling law suggests the side jump mechanism responsible for the observed AHE. The large value of the exponent n = 2.2, indicates the role of interfaces in the enhancement of AHE. The Hall sensitivity of 1.2 Ω/T and 0.9 Ω/T was observed for the samples with $t_{Fe}$=10 Å and 20 Å respectively, which were about three order of magnitudes larger than that of Al/Fe and Cu/Fe multilayers. Our AHE data shows that Si/Fe multilayers may be a possible candidate for Hall element for potential applications in the field of magnetic sensors.

# References

[1] Baibich M. N., Broto J. M., Fert A., Van Dau F. N., Petroff F., Etienne P., Creuzet G., Friederich A. and Chazelas J. 1988 *Phys. Rev. Lett.* **61** 2472-5.

[2] Binasch G., Grünberg P., Saurenbach F. and Zinn W. 1989 *Phys. Rev. B* **39** 4828-30.

[3] Patrin G. and Vas'kovskii V. 2006 *Phys. Met. Metallogr.* **101** S63-S66.

[4] Gerber A. and Riss O. 2008 *J. Nanoelectron. Optoelectron.* **3** 35-43.

[5] Kumar S. and Laughlin D. E. 2005 *IEEE Trans. Magn.* **41** 1200-8.

[6] Lu Y. M., Cai J. W., Pan H. Y. and Sun L. 2012 *Appl. Phys. Lett.* **100** 022404.

[7] Song S. N., Sellers C. and Ketterson J. B. 1991 *Appl. Phys. Lett.* **59** 479-81.

[8] Khatua P., Majumdar A. K., Temple D. and Pace C. 2006 *Phys. Rev. B* **73** 094421.

[9] Liu Y. W., Mi W. B., Jiang E. Y. and Bai H. L. 2007 *J. Appl. Phys.* **102** 063712-7.

[10] Zhu Y. and Cai J. W. 2007 *Appl. Phys. Lett.* **90** 012104-13.

[11] Chien C. L. and Westgate C. R. 1980 *The Hall Effect and its Applications* (New York: Plenum Press)

[12] Liu H., Zheng R. K. and Zhang X. X. 2005 *J. Appl. Phys.* **98** 086105.

[13] O'Handley R. C. 1999 *Modern Magnetic Materials Principles and Applications* (New York: Wiley)

[14] Hurd C. M. 1972 *The Hall Effect in Metals and Alloys* (New York: Plenum Press)

[15] Volkov V., Levashov V., Matveev V., Matveeva L., Khodos I. and Kasumov Y. 2011 *Thin Solid Films* **519** 4329-33.

[16] Galepov P. S. 1969 *Soviet Physics Journal* **12** 133-5.

[17] Xu W., Zhang B., Wang Z., Chu S., Li W., Wu Z., Yu R. and Zhang X. 2008 *The Eur. Phys. J. B - Condensed Matter and Complex Systems* **65** 233-7.

[18] Zhao B. and Yan X. 1997 *J. Appl. Phys.* **81** 4290-2.

[19] Aronzon B., Kovalev D., Lagar'kov A., Meilikhov E., Ryl'kov V., Sedova M., Negre N., Goiran M. and Leotin J. 1999 *JETP Lett.* **70** 90-6.

[20] William D. Callister J. 2001 *Fundamentals of Materials Science and Engineering, Fifth edition* (New York: John Wiley & Sons, Inc.) pp 377.

[21] Das S. S. and Kumar M. S. 2013 *AIP Conf. Proc.* **1512** 666-7.

[22] Grundy P. J., Fallon J. M. and Blythe H. J. 2000 *Phys. Rev. B* **62** 9566-74.

[23] Brajpuriya R., Sharma R., Vij A. and Shripathi T. 2011 *Journal of Modern Physics* **2** 864-74.

[24] Cullity B. D. and Graham C. D. 2009 *Introduction to Magnetic Materials* (Piscataway: IEEE Press Willey)

[25] Denardin J. C., Pakhomov A. B., Knobel M., Liu H. and Zhang X. X. 2001 *J. Magn. Magn. Mater.* **226-230, Part 1** 680-2.

[26] Pakhomov A. B., Yan X. and Xu Y. 1996 *J. Appl. Phys.* **79** 6140-2.

[27] Polley C., Clarke W. and Simmons M. 2011 *Nanoscale Res. Lett.* **6** 538.

[28] Murarka S. P., Gutmann R. J., Kaloyeros A. E. and Lanford W. A. 1993 *Thin Solid Films* **236** 257-66.

[29] Karplus R. and Luttinger J. M. 1954 *Phys. Rev.* **95** 1154-60.

[30] Smit J. 1958 *Physica* **24** 39-51.

[31] Berger L. 1970 *Phys. Rev. B* **2** 4559-66.

[32] Nagaosa N., Sinova J., Onoda S., MacDonald A. H. and Ong N. P. 2010 *Rev. Mod. Phys.* **82** 1539-92.

[33] Tsui F., Chen B., Barlett D., Clarke R. and Uher C. 1994 *Phys. Rev. Lett.* **72** 740-3.

[34] Guo Z. B., Mi W. B., Aboljadayel R. O., Zhang B., Zhang Q., Barba P. G., Manchon A. and Zhang X. X. *Phys. Rev. B* **86** 104433.

[35] Zhang S. 1995 *Phys. Rev. B* **51** 3632-6.